\documentclass[a4paper,amsmath,amssymb]{jpconf}
                                                                                
\usepackage{iwsmihead}
\usepackage{graphicx} 
\usepackage{amsmath, amsthm, amssymb}

\begin{document}
\def\be{\begin{equation}}
\def\ee{\end{equation}}
\def\bc{\begin{center}}
\def\ec{\end{center}}                                                                                
\title{A Novel Quantum Transition in a Fully Frustrated Transverse Ising Antiferromagnet }
                                                                                
\author{Anjan Kumar Chandra}

\address{%
Centre for Applied Mathematics and Computational Science, Saha Institute of Nuclear Physics, 1/AF Bidhannagar, Kolkata-700064, India.
}%

\ead{anjan.chandra@saha.ac.in}

\author{Jun-ichi Inoue}%

\address{%
Complex Systems Engineering, Graduate School of Information
Science and Technology, Hokkaido University, 
N14-W9, Kita-ku, Sapporo 060-0814, Japan\\
}%

\ead{j$\underline{\,\,\,}$inoue@complex.eng.hokudai.ac.jp}

\author{Bikas K. Chakrabarti}%

\address{%
Centre for Applied Mathematics and Computational Science and 
Theoretical Condensed Matter Physics Division, Saha Institute of Nuclear Physics, 1/AF Bidhannagar, Kolkata-700064, India.\\
}%

\ead{bikask.chakrabarti@saha.ac.in}

\date{\today}

\begin{abstract}
We consider a long-range Ising antiferromagnet (LRIAF) put in a transverse field. Applying quantum
Monte Carlo method, we study the variation of order parameter (spin correlation in Trotter time
direction), susceptibility and average energy of the system for various values of the transverse
field at different temperatures. The antiferromagnetic order is seen to get immediately broken as
soon as the thermal or quantum fluctuations are added. We also discuss the phase diagram for 
the Sherrington-Kirkpatrick (SK) model with the same LRIAF bias, also in presence of a transverse
field. We find that while the antiferromagnetic order is immediately broken as one adds an
infinitesimal transverse field or thermal fluctuation to the system, an infinitesimal SK spin glass disorder is enough to induce a stable glass order in the antiferromagnet. This glass order 
eventually gets destroyed as the thermal or quantum fluctuations increased beyond their threshold values and the transition to para phase occurs.  Indications of this novel phase transition are discussed. Because of the presence of full frustration, this  surrogate property of the LRIAF for incubation of stable spin glass phase in it (induced by addition of a small disorder) should  enable eventually the study of classical and quantum spin glass phases by using some perturbation 
theory with respect to the disorder. 
\end{abstract}

\def\be{\begin{equation}}
\def\ee{\end{equation}}

\section{Introduction} 
Quantum phases in frustrated systems are being intensively investigated these
days; in particular in the context of quantum spin glass and quantum ANNNI
models \cite{Bhatt}. Here we study a fully-frustrated quantum antiferromagnetic
model. Specifically, the long-range antiferromagnetic Ising model put under
transverse field. The finite temperature properties of sub-lattice decomposed 
version of this model was already
considered earlier \cite{Chakrabarti,Das}. The quantum phase transition and 
entanglement properties of the full long-range model at zero temperature was 
studied by Vidal et al \cite{Vidal}. Here we 
present some results obtained by applying quantum Monte Carlo
technique \cite{Suzuki} to the same full long-range model at finite 
temperature. We observe indications of a
quantum phase transition in the model, where the 
antiferromagnetically 
ordered phase gets  
destabilised by both 
infinitesimal thermal (classical) as well quantum 
fluctuations (due to tunneling or transverse field) and the system becomes
disordered or goes over to the para phase.

The ordered phase of the long range Ising Antiferromagnet (LRIAF) seems to be
extremely volatile and loses the order (freezing of spin orientations) at
any finite fluctuation level; classical or quantum. However the LRIAF model
has the required frustration of the Sherrington-Kirkpatrick (SK)  model, which
could support the spin glass order, but for any
disorder. To check if this `liquid'-like antiferromagnetic phase
of LRIAF can get `crystallized' into spin-glass phase if a little disorder is
added, we study next this LRIAF Hamiltonian with a tunable coupling with the SK spin
glass Hamiltonian and study this entire system's phase transitions induced by both thermal
and tunneling field. Indeed, the stable SK spin glass phase is observed for thermal
or quantum fluctuations below finite threshold values.  

We employ the analytic 
(mean field) solution 
of  the transverse Ising model 
with long-range interactions can be in presence of the special kind of
quenched disorder appropriate for 
the SK 
spin glasses. 
With this,  
we  study  the phase diagram for 
the SK model 
with LRIAF  bias 
in a transverse field. 
We again find that 
the antiferromagnetic order 
is immediately broken 
when one adds an infinitesimal 
transverse field or thermal fluctuation 
to the system. However,  an infinitesimal SK-type
disorder is enough to make the system `crystallized' into a glass phase.

This paper is organized in the following manner.
In Section 2, we introduce the pure quantum LRIAF model and then 
discuss the
(finite temperature)  quantum Monte Carlo
results. 
In Section 3, 
we consider 
the SK model 
with antiferromagnetic bias 
in a transverse field, and discuss the (analytic) mean field phase diagram. 
In Section 4, we present some discussions on our results.
\section{The pure LRIAF  model}
The Hamiltonian of the infinite-range quantum Ising 
antiferromagnet (without any spin glass disorder) is 
\begin{eqnarray}
H & \equiv & H^{(C)} + H^{(T)} = \frac{J_0}{N}\sum_{i,j(>i)=1}^N\sigma^z_i \sigma^z_j  
-  h\sum_{i=1}^N \sigma^z_i - \Gamma \sum_{i=1}^N \sigma^x_i ,
\label{eq:Hamiltonian1}
\end{eqnarray}
where $J_0$ denotes the long-range antiferromagnetic $(J_0>0)$ exchange constant;
for convenience,
we fix the value $J_0 = 1$ in this section.
Here $\sigma^x$ and $\sigma^z$ denote the $x$ and $z$ component of the $N$ 
Pauli spins

\[ \sigma^{z}_i = \left( \begin{array}{cc}
1 & 0 \\
0 & -1 \\ \end{array} \right); \hspace*{0.5cm}
\sigma^{x}_i = \left( \begin{array}{cc}
0 & 1 \\
1 & 0 \\ \end{array} \right); \hspace*{0.5cm} i = 1,2,....,N.
\]
$h$ and $\Gamma$ denote respectively the longitudinal and transverse 
fields. We have denoted the co-operative term of 
{$H$} (including the external longitudinal field term) by $H^{(C)}$ and the 
transverse field part as $H^{(T)}$.
As such the model has a fully frustrated (infinite-range or infinite dimensional) co-operative term. At zero temperature and at zero longitudinal and 
transverse fields, the 
$H^{(C)}$ would prefer the spins to orient in $\pm z$ directions only with
zero net magnetization in the $z$-direction. This antiferromagnetically 
ordered state is 
completely frustrated and highly degenerate. Switching on the transverse field
$\Gamma$ would immediately induce all the spins to orient in the
$x$-direction (losing the degeneracy), corresponding to a maximum of the 
kinetic energy term and this discontinuous
transition to the para phase occurs at $\Gamma = 0$. However, at any
finite temperature the entropy term coming from the extreme degeneracy of the
antiferromagnetically ordered state and the close-by excited states does not
seem to  
induce a stability of this phase. 
\begin{figure}
\bc
\noindent \includegraphics[clip,width= 9cm, angle = 0]
{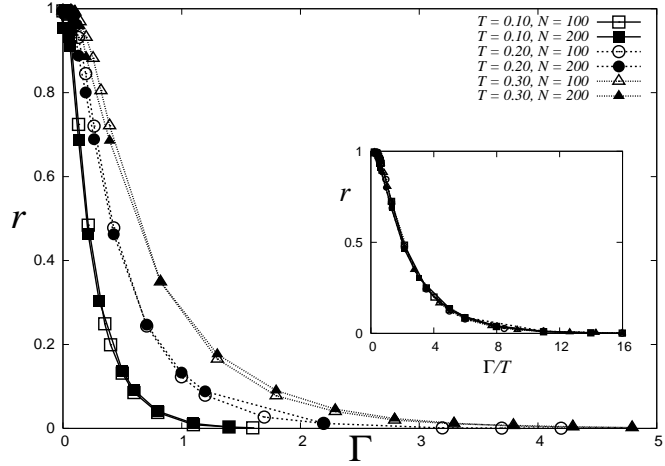}
\caption{\footnotesize 
\label{fig:insetq}Variation of the order parameter $r$ 
(correlation in the Trotter direction)
with transverse field $\Gamma$ for $T = 0.10, 0.20$ and 
$0.30$ ($h = 0$) for two different system sizes ($N = 100$ and $200$). 
$r = 0$ for large  $\Gamma $.
 The inset shows the plot of $r$ against the scaled variable $\Gamma/T$.}
\ec
\end{figure}
\begin{figure}
\bc
\noindent \includegraphics[clip,width= 9cm, angle = 0]
{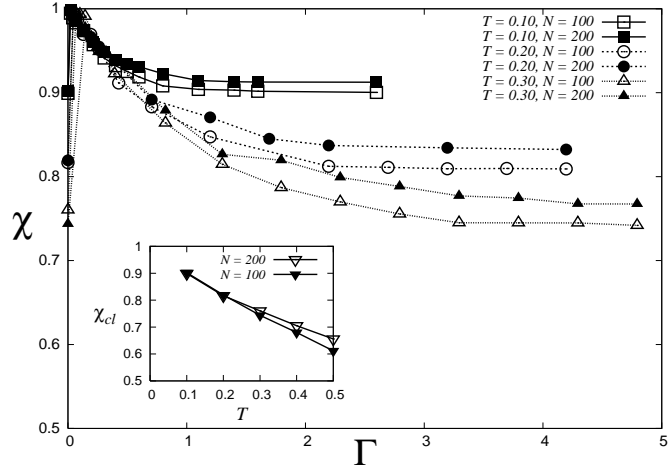}
\caption{\footnotesize 
\label{fig:inset5}Variation of the 
susceptibility $\chi$ with transverse field $\Gamma$ for
$T = 0.10, 0.20$ and $0.30$ ($h \le 0.1$) for two different system sizes 
($N = 100$ and $200$). The corresponding susceptibility $\chi_{cl}$ for various
temperatures for $N = 100$ and $200$ for the
classical system are shown in the inset. $\chi$ 
converges to the classical values $\chi_{cl}$ for 
large $\Gamma$.}
\ec
\end{figure}
\begin{figure}
\bc
\noindent \includegraphics[clip,width= 9cm, angle = 0]{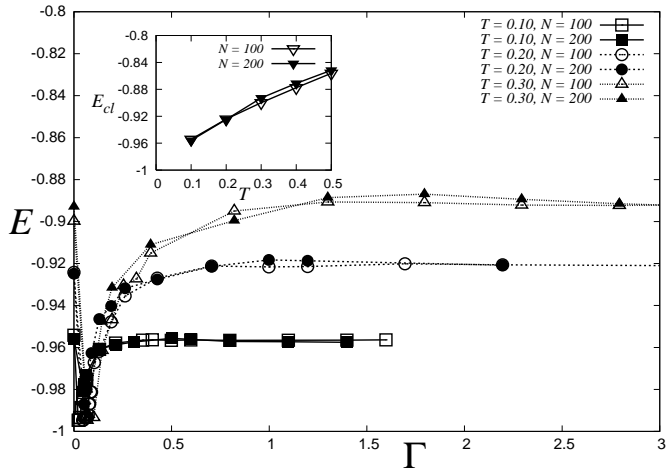}
\caption{\footnotesize 
\label{fig:inset4}Variation of average energy $E$ with 
transverse field $\Gamma$ for $T = 0.10, 0.20$ and $0.30$ ($h = 0$) for 
two different values 
of $N$($=100,200$). The corresponding average energy $E_{cl}$ for various 
temperatures for $N = 100$ and $200$ for the are shown in the inset. 
$E$ converges to the classical values $E_{cl}$ for large $\Gamma$ .  }
\ec
\end{figure}
\subsection{Monte Carlo simulation}
\subsubsection{Suzuki-Trotter mapping and simulation}
This Hamiltonian (\ref{eq:Hamiltonian1}) can be mapped to a 
$(\infty + 1)$-dimensional classical 
Hamiltonian \cite{Suzuki}
using the Suzuki-Trotter formula. The effective Hamiltonian 
can be written as 
\begin{eqnarray}
{\mathcal H} & = & \frac{1}{NP} \sum_{i,j(>i)=1}^N 
\sum_{k=1}^P \sigma_{i,k} \sigma_{j,k} \notag - \frac{h}{P} 
\sum_{i=1}^N \sum_{k=1}^P \sigma_{i,k} - \frac{J_p}{P} \sum_{i=1}^N 
\sum_{k=1}^P \sigma_{i,k} \sigma_{i,k+1},
\end{eqnarray}
where
\begin{equation}
J_p = -(PT/2)\ln(\tanh(\Gamma/PT)) .
\end{equation}
Here $P$ is the number of Trotter replicas and $k$ denotes the 
$k$-th row in
the Trotter direction. $J_p$ denotes the nearest-neighbor interaction 
strength along the Trotter direction. We have studied the system for $N = 100$.
Because of the diverging growth of interaction $J_p$ for very low values of 
$\Gamma$ and also for high values of $P$, and the consequent non-ergodicity 
(the system relaxes to different states for identical thermal and quantum parameters, due to frustrations, starting from 
different initial configurations), we have kept
the value of $P$ at a fixed value of $5$. 
This choice of $P$ value helped satisfying the ergodicity 
of the system up to
very low values of the transverse field at the different 
temperatures considered
$T = 0.10$ and $0.20$.
Starting from a random initial configurations 
(including all up or 50-50 
up-down configurations) we 
follow the time variations of different 
quantities until they relax and
study the various quantities after they relax. 
\subsubsection{Results}
We studied results for three different temperatures 
$T = 0.10, 0.20$ and $0.30$ and all the results 
are for $N = 100$ and $200$ and $P = 5$.
We estimated the following quantities after relaxation :

\medskip
(i) {\em Correlation along Trotter direction ($r$)} : 
We studied the variation
of the order parameter 
\begin{equation}
r = \frac {1}{NP}\sum_{i=1}^N \sum_{k=1}^P \langle \sigma_{i,k} \sigma_{i,k+1}  \rangle,
\end{equation}
which is the first neighbor correlation along Trotter direction. Here, 
$\langle ... \rangle$ indicate the average over initial
spin configurations. This quantity $r$ shows a smooth vanishing behavior. 
We consider this correlation $q$ as the order parameter for the transition at 
$\Gamma_c$. A larger transverse field is needed for the vanishing of the
order parameter for larger temperature. The observed values 
(see Figure \ref{fig:insetq}) of $\Gamma_c$ 
are $\simeq 1.6, 2.2$ and $3.0$ for $T = 0.1, 0.2$ and $0.3$ respectively.
As shown in the inset, an unique data collapse occurs when $r$ is plotted
against $\Gamma/T$ and one seems to get the complete disorder immediately as
the scaling dos not involve any finite value $T_c$. This is consistent
with the observations in the next section.

\medskip

(ii) {\em Susceptibility ($\chi$)} : The longitudinal
susceptibility $\chi = (1/NP) {\partial [\sum_{i,k} 
\langle\sigma_{i,k}\rangle}]/{\partial h}$,
where $h$ ($\rightarrow 0$) is the applied longitudinal field, has also been
measured. We went up to $h = 0.1$ and estimated the $\chi$ values. As we 
increase the value of the 
transverse field $\Gamma$ from a suitably chosen low value, $\chi$ initially 
starts with a value almost equal to unity and then gradually saturates at lower
values (corresponding to the classical system where $J_p = 0$ in Eq.(2)) as 
$\Gamma$ is increased. Also at $\Gamma = 0$, the classical values are 
indicated in Figure \ref{fig:inset5}. This 
saturation value of $\chi$ decreases with temperature. Again the field at
which the susceptibility saturates are the same as for the vanishing of the
order parameter for each temperature.
\medskip 

(iii) {\em Average energy (E)} : We have measured the value of 
the
co-operative energy for each Trotter index and then take its average $E$ i.e.
$E = \langle H^{(C)} \rangle$ of Eq. (1) with $J_0 = 1$. It initially begins
with $-1.0$ and after a sharp rise the average energy saturates,
at large values of $\Gamma$, to
values corresponding to the classical equilibrium energy ($E_{cl}$ for $J_p = 0$ 
in Eq.(2)) at those
temperatures. Again it takes larger values of $\Gamma$ at higher temperatures 
to achieve the classical equilibrium energy. At $\Gamma = 0$, the corresponding
classical values of $E$ are plotted in Figure \ref{fig:inset4}.
The variations of all these quantities indicate that the `quantum order' 
disappears
and the quantities reduce to their classical values 
(corresponding to $J_p = 0$ in for large values of  
the transverse field $\Gamma$.
\subsection{A mean field analysis}
First, Let us consider the case of pure LRIAF model and
  rewrite our Hamiltonian
{$H$} in Eq.(1) for 
$\tilde{J}=0$ as 
\begin{equation}
H =  \frac{1}{2N} \left(
{ \sum_{i=1}^N \sigma^z_i}
\right)^2  -  \frac{1}{N} \sum_{i=1}^N {(\sigma^z_i)}^2 - h \sum_{i=1}^N \sigma^z_i - \Gamma \sum_{i=1}^N \sigma^x_i 
\end{equation}
If we now denote the total spin by $\vec{\sigma}_{tot}$ i.e.
$\vec{\sigma}_{tot} = \frac{1}{N} \sum_{i=1}^N \vec{\sigma}_i$ 
(where $N|\vec{\sigma}|= 0,1,2,....,N$), then the Hamiltonian 
{$H$} can be expressed as 
\begin{equation}
\frac{H}{N} =  \frac{1}{2} {(\sigma^z_{tot})}^2 - h \sigma^z_{tot} - \Gamma \sigma^x_{tot} - \frac{1}{N} .
\end{equation}
Let us assume the average total spin $\langle\vec{\sigma}\rangle$ to be oriented at an angle $\theta$ with
the $z$-direction : $\langle\sigma^z_{tot}\rangle = m \cos \theta$ and 
$\langle\sigma^x_{tot}\rangle = m \sin \theta$. Hence the average 
total energy $E_{tot} = \langle H \rangle$ can be written as
\begin{equation}
\frac{E_{tot}}{N} =  \frac{1}{2} m^2 {{\cos}^2 \theta} - h m {\cos \theta} - \Gamma m  \sin \theta - \frac{1}{N} .
\end{equation}
At the zero temperature and at $\Gamma = 0$, for $h = 0$, the energy 
$E_{tot}$ is 
minimised when {$\theta = 0$} and {$m = 0$} (complete antiferromagnetic
order in $z$-direction). As soon as $\Gamma \ne 0$ ($h = 0$) the minimisation
of $E_{tot}$ requires {$\theta = \pi/2$} and $m = 1$ (the maximum possible
value); driving the system to paramagnetic phase. This discontinuous transition
at $T = 0$ was also seen in \cite{Vidal}. 
As observed in our Monte Carlo study in the previous 
section, $\Gamma_c(T) \rightarrow 0$ as 
$T \rightarrow 0$. This is consistent with this exact result $\Gamma_c = 0$
at $T = 0$. For $T = 0$ (and $h = 0$), therefore, the transition from
antiferromagnetic ($\theta = 0 = m$) to para ($\theta = \pi/2, m = 1$)
phase, driven by the transverse field $\Gamma$, occurs at $\Gamma = 0$ itself.

One can also
estimate the susceptibility $\chi$ at $\Gamma = 0 = T$. Here 
$E_{tot}/N = \frac{1}{2} m^2 {{\cos}^2 \theta} - h m {\cos \theta} - \frac{1}{N}$ and the minimisation of
this energy gives $m \cos \theta = h$ giving the (longitudinal) 
susceptibility $\chi = {m \cos \theta}/{h} = 1$. This is consistent with
the observed behaviour of $\chi$ shown in Figure \ref{fig:inset5} where the 
extrapolated value of
$\chi$ at $\Gamma = 0$ increases with decreasing $T$ and approaches $\chi = 1$
as $T \rightarrow 0$. 

At finite temperatures $T \ne 0$, for $h = 0$, we have to consider also the 
entropy term 
and minimise the free energy ${\mathcal F} = E_{tot} - TS$ rather than 
$E_{tot}$ where $S$ denotes the entropy of the state.
This entropy term will also take part in fixing the value of $\theta$ and 
$m$ at which the free energy ${\mathcal F}$ is minimised. As soon
as the temperature $T$ becomes non-zero, the extensive entropy of the system 
for 
antiferromagnetically ordered state with $m \simeq 0$ (around and close-by
excited states with $\theta = 0$) helps stabilisation 
near $\theta = 0$ and $m = 0$ rather than near the para phase with
$\theta = \pi/2$ and $m = 1$, where the entropy drops to zero. While the 
transverse field tends to align the spins along $x$
direction (inducing $\theta = \pi/2$ and $m = 1$), the entropy factor 
prohibits that and the system adjusts $\theta$ and
$m$ values accordingly and they do not take the disordered or para state 
values
($\theta = \pi/2$ and $m = 1$) for any non-zero value of $\Gamma$ (like
at $T = 0$).
For very large values of $\Gamma$, of
course, the free energy ${\mathcal F}$ is practically dominated by the
transverse field term in $H$ and again {$\theta = \pi/2$}
and $m = 1$, beyond $\Gamma = \Gamma_c(T) > 0$ for $T > 0$.
However, this continuous transition-like behaviour may be argued 
\cite{Diptiman} to correspond to a 
crossover type property of the model at finite temperatures (suggesting
that the observed finite values of $\Gamma_c(T)$ are only effective numerical
values). In fact, for $h = 0$ one adds the entropy term $-T\ln D_s$ to
$E_{tot}$ in Eq.(7) to get ${\mathcal F}$ and 
one can then get \cite{Diptiman}, after minimising the ${\mathcal F}$ with
respect to $m$ and $\theta$, $m = \tanh (\Gamma/2T)$, which 
indicates an analytic variation of $m$ 
and no phase transition 
at any finite temperature for $\tilde{J}=0$
(antiferromagnetic phase 
occurs only at 
$\Gamma=T=0$ as shown in 
Figure \ref{fig:fg2}. 
\section{LRIAF with SK disorder: `Liquid' phase of the SK spin glasses gets frozen}
In the last part of this paper, we discuss the phase diagram for 
the Sherrington-Kirkpatrick model 
with antiferromagnetic bias 
in a transverse field. 
We find that 
the antiferromagnetic order 
is immediately broken 
when one adds an infinitesimal 
transverse field or thermal fluctuation 
to the system, 
whereas an infinitesimal SK-type 
disorder is enough to get the system `crystalized' into the glass phase.

The model we discuss here is 
given by the following Hamiltonian
\begin{eqnarray}
H & = & \frac{1}{N}
\sum_{ij (j > i) }
(J_{0}-\tilde{J}\tau_{ij})
\sigma_{i}^{z}
\sigma_{j}^{z}-
\Gamma \sum_{i}\sigma_{i}^{x}
\label{eq:Hamiltonian}
\end{eqnarray}
where 
$J_{0}$ is a parameter 
whcih controlls 
the strength of the antiferromagnetic 
bias and 
$\tilde{J}$ is 
an amplitude of the disorder 
$\tau_{ij}$ in each pair interaction. 
The $\Gamma$ controls the quantum-mechanical fluctuation. 
When we assume that 
the disorder 
$\tau_{ij}$ obeys 
a Gaussian with mean zero and 
variance unity, 
the new variable $J_{ij} 
\equiv 
-J_{0}+ 
\tilde{J}\tau_{ij}$ follows the following distribution. 
$P(J_{ij}) =  
{\exp}[
-{(J_{ij}+J_{0})^{2}}
/{2\tilde{J}^{2}}]/{\sqrt{2\pi}\tilde{J}}$. 
Therefore, 
we obtain the `pure' antiferromagnetic 
Ising model with infinite range interactions 
when we consider the limit 
$\tilde{J} \to 0$ keeping 
$J_{0} > 0$. 
On ther other hand, 
for $J_{0} < 0$ with 
$\Gamma =0$ is 
identical to the classical SK model. 
In this paper, 
we investigate the condition 
for which the antiferromagnetic order survives. 

For the Hamiltonian (\ref{eq:Hamiltonian}), 
we immediately obtain 
the saddle point equations 
under the static and 
replica symmetric approximations as follows (see e.g. 
\cite{CDS,RMP-Das}). 
\begin{eqnarray}
m & = & 
\int_{-\infty}^{\infty}
Dz 
\frac{(\tilde{J}\sqrt{q}z +  K_{0}m)}
{\sqrt{(\tilde{J}\sqrt{q}z + K_{0}m)^{2}+
\Gamma^{2}}}
\tanh \beta 
\sqrt{(\tilde{J}\sqrt{q}z + K_{0}m)^{2}+
\Gamma^{2}} 
\label{eq:quantum_m} \\
q & = &  
\int_{-\infty}^{\infty}
Dz 
\left\{
\frac{(\tilde{J}\sqrt{q}z + K_{0}m)}
{\sqrt{(\tilde{J}\sqrt{q}z + K_{0}m)^{2}+
\Gamma^{2}}}
\right\}^{2}
\tanh^{2} \beta 
\sqrt{(\tilde{J}\sqrt{q}z + K_{0}m)^{2}+
\Gamma^{2},}
\label{eq:quantum_q} 
\end{eqnarray}
where $K_{0} \equiv {\rm sgn}(J_{0})|J_{0}|$ and 
$m \equiv N^{-1}
\sum_{i}
\sigma_{i}^{z}$ is a 
magnetization 
and 
$q \equiv N^{-1}
\sum_{i}
\langle \sigma_{i}^{z} \rangle^{2}$ 
is a spin glass order parameter. 
We defined 
$Dz \equiv dz\,
{\rm e}^{-z^{2}/2}/
\sqrt{2\pi}$. 
The bracket $\langle \cdots \rangle$ denotes 
an expectation 
over the density matrix: 
$\rho =
{\rm e}^{-\beta H}/{\rm tr}
\,{\rm e}^{-\beta H}$. When $J_0$ is negative, (\ref {eq:quantum_m}) has the
only solution $m =0$.  The general phase boundaries (see Figure 4) between the
ferro (F), spin glass (SG), antiferro (AF) and para (P) phases  can be obtained by 
solving the 
the above two equations in the limit $m \rightarrow o; q \ne 0$ for the F-SG
boundary, $q \rightarrow 0$ for SG-P boundary and $m \rightarrow 0$ fot the
F-P boundary.  In these limits, the P-SG boundary equations become (see
e.g. \cite {RMP-Das})
\begin{equation}
\Gamma = \tilde{ J}\tanh \left ( \Gamma \over T\right)
\label{eq:boundary}.
\end{equation}
 
\subsection{Classical system}
In the classical limit, the equations of state 
are simplified as $m =   
\int_{-\infty}^{\infty}
Dz 
\tanh 
\beta 
(\tilde{J}\sqrt{q}z -J_{0}m)$, 
$q = 
\int_{-\infty}^{\infty}
Dz 
\tanh^{2}
\beta 
(\tilde{J}\sqrt{q}z -J_{0}m)$. 
For $J_{0} > 0$, 
we find that $m=0$ is 
only physical solution 
for all temperature regimes. 
This fact means that 
there are three possible phases, 
namely, the antiferromagnetic phase, 
the paramagnetic 
phase and 
the spin glass phase. 
In these three phases, 
the magnetization $m$ is zero. 
To determine the 
critical point $T_{SG}$ 
at which the spin glass transition 
takes place, 
we expand the equation 
with respect to $q$ for $q \simeq 0$ and $m=0$. 
Then, we have 
$T_{SG} = \tilde{J}$ 
and the critical point 
is independent of 
the antiferromagnetic bias $J_{0}$. 
\begin{figure}[ht]
\begin{center}
\includegraphics[width=10cm]{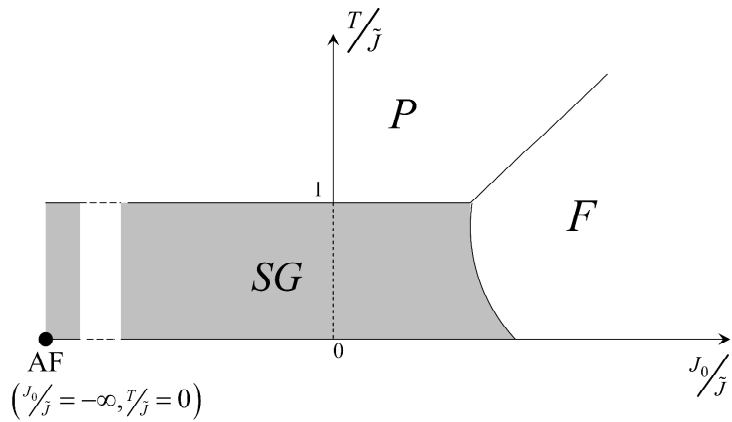}
\end{center}
\caption{\footnotesize 
The phase diagram of 
classical SK model \cite{Binder} extended for antiferromagnetic bias. 
For $J_{0}<0$, 
there exist spin glass phase 
below $T/\tilde{J}=1$ and 
the critical temperature is 
independent of the strength of the 
antiferromagnetic bias $J_{0}$. 
For finite temperature $T>0$, 
the anti-ferromagnetic order dissapears and 
the system changes to the paramagnetic phase. 
When we add an infinitesimal disorder 
$\tilde{J}>0$, 
the antiferromagnetic order 
is broken down and 
the system suddenly gets `crystalized' into 
a spin glass (SG) phase.}
\label{fig:fg2}
\end{figure}
\mbox{} 
This result means 
that 
the antiferromagnetic order 
can apprear if 
and only if 
we set $J_{0}>0$ and $T/\tilde{J}=0,J_{0}/\tilde{J}=0$. 
On the other hand, 
for $0 < J_{0} < \infty$ at low temperature regime 
$T < T_{SG}$, the spin glass phase appears. 
We plot the phase diagram in Figure \ref{fig:fg2}. 
We also 
conclude that 
the system 
described by the Hamiltonian 
(\ref{eq:Hamiltonian}) with 
$\Gamma =0$ is immediately 
`crysterlized' when 
we add any 
infinitesimal 
disorder $\tilde{J} >0$. 

From the view point of 
the degeneracy of 
the spin configurations, 
we easily estimate the 
number of 
solution for 
the antiferromagnetic 
phase as $2^{N/2}={\rm e}^{0.346N}$, 
which is larger than 
the number of the SK model ${\rm e}^{0.199N}$ \cite{Binder}. 
However, 
for the infinite range antiferromagnetic 
model, the energy barrier between 
arbitrary configurations which gives the same 
lowest energy states is of order $1$ and there is no ergodicity breaking.  
\subsection{Quantum system}
We next consider the case of 
presence of transverse field 
$\Gamma \neq 0$. 
In this case, we also find that 
the saddle point equation 
(\ref{eq:quantum_m}) has 
a solution 
$m=0$ and 
the phase boundary 
between 
the spin glass and 
paramagnetic phases is given by 
setting $m=0$ and $q \simeq 0$ and 
we get (\ref{eq:boundary}). 
Obviously, the boundary 
at $T=0$ gives 
$\Gamma_{SG}=\tilde{J}$. 
On the other hand, 
when we consider the case of 
$\Gamma \simeq 0$, we have 
$T_{SG}=\tilde{J}$. 
These fact means 
that there is no antiferromagnetic nor 
the spin glass phase 
when we consider the pure case 
$\tilde{J}=0$ because 
the critical point leads to 
$T_{SG}=\Gamma_{SG}=0$. 
Therefore, 
we conclude that 
the antiferromagnetic phase 
can exisit 
if and only if 
$T=\Gamma=0$. 
\begin{figure}[ht]
\begin{center}
\includegraphics[width=10cm]{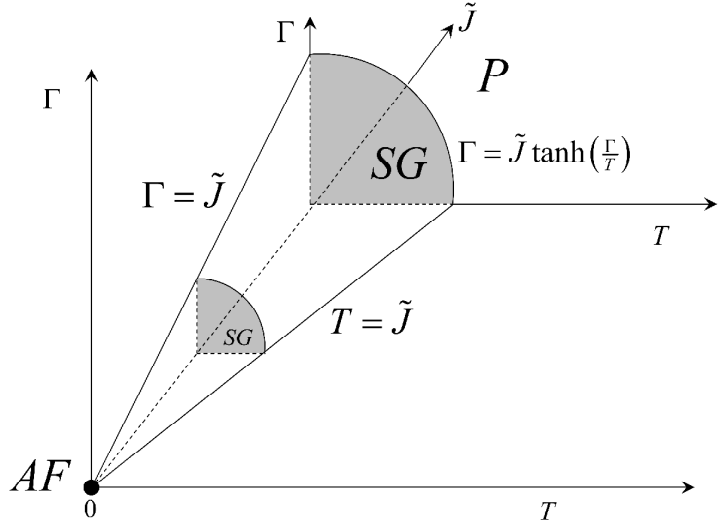}
\end{center}
\caption{\footnotesize 
Phase diagram for 
the quantum system. 
The antiferromagnetic order 
exisits if and only if we set 
$T=\Gamma=0$. 
As the $\tilde{J}$ decreases, 
the spin glass phase gradually shrinks to 
zero 
and eventually ends up 
at an antiferromagnetic phase 
at its vertex (for $\Gamma=0=T=\tilde{J}$) as 
discussed in Section 2. 
}
\end{figure}
\section{Discussion}

We considered here a long-range Ising antiferromagnet at a 
finite temperature and put in a transverse field. The antiferromagnetic
order is seen to get immediately broken as soon as the thermal or
quantum fluctuations are added.  
However, when we add the Sherrington-Kirkpatrick hamiltonian as
perturbation we find that an infinitesimal 
SK spin glass disorder is enough to 
induce a stable glass order in this LRIAF antferromagnet. This glass order 
eventually gets destroyed as
the thermal or quantum fluctuations increased beyond their threshold values and
the transition to para phase occurs. 
As shown in the phase diagram in Figure 5,  the antiferromagnetic phase of the LRIAF (ocurring
only at $\tilde{J} = 0 = \Gamma = T$), can get `crystalised' into 
spin-glas phase if 
a little SK-type disoder is
added ($\tilde{J} \neq 0$); the only missing element in the LRIAF 
(which is fully frustrated, but lacks disorder)
to induce stable order 
(freezing of random soin orientations) in it.
As mentioned already, 
the degeneracy factor ${\rm e}^{0.346N}$ 
of the ground state of the LRIAF is 
much larger than that ${\rm e}^{0.199N}$ for the 
SK model.  
Hence, (because of the presence of 
full frustration) the LRIAF posses a surrogate incubation property of
stable spin glass phase in it when induced by addition of a small disorder.
 This observation
should  enable eventually the
study of 
classical and quantum spin glass phases 
by using some perturbation 
theory with respect to the disorder.

\section*{Acknowledgements}
The work of one author (AKC) was supported by the Centre for Applied
Mathematics and Computational Science (CAMCS) of the Saha Institute of
Nuclear Physics. We are grateful to I. Bose, A. Das, S. Dasgupta, D. Sen, 
P. Sen and K. Sengupta for
useful discussions and comments. 
One of the authors (BKC) 
thanks Hokkaido University 
for hospitality. 
Researches at Hokkaido University were financially supported 
by {\it Grant-in-Aid 
Scientific Research on Priority Areas 
``Deepening and Expansion of Statistical Mechanical Informatics (DEX-SMI)" 
of The Ministry of Education, Culture, 
Sports, Science and Technology (MEXT)} 
No. 18079001.

\end{document}